\documentclass[twocolumn,showpacs]{revtex4}
\newcommand{\gtsim}{\mbox{{\raisebox{-0.4ex}{$\stackrel{>}{{\scriptstyle\sim}}
$}}}}
\newcommand{\ltsim}{\mbox{{\raisebox{-0.4ex}{$\stackrel{<}{{\scriptstyle\sim}}
$}}}}

\usepackage{graphicx}
\usepackage{dcolumn}
\usepackage{amsmath}
\usepackage{longtable}
\begin{document}

\title{Experimental electronic heat 
capacities of $\alpha-$ and $\delta-$Plutonium;
heavy-fermion physics in an element}

\author{J.C.~Lashley$^1$, J.~Singleton$^1$, A.~Migliori$^1$,
J.B.~Betts$^1$, R. A.~Fisher$^2$, J. L.~Smith$^1$, and
R.J.~McQueeney$^1$}

\affiliation{$^1$MST-Division, Los Alamos National Laboratory,
Los Alamos, NM-87545, USA\\
$^2$Lawrence Berkeley National Laboratory, 
Berkeley, CA94720, USA}

\begin{abstract}
We have measured the heat capacities of
$\delta-$Pu$_{0.95}$Al$_{0.05}$ and $\alpha-$Pu over the
temperature range $2-303$~K. The availability of data below 10~K
plus an estimate of the phonon contribution to the heat capacity
based on recent neutron-scattering experiments 
on the same sample enable
us to make a reliable deduction of the electronic
contribution to the heat capacity of
$\delta-$Pu$_{0.95}$Al$_{0.05}$; we find
$\gamma = 64 \pm 3$~mJK$^{-2}$mol$^{-1}$
as $T \rightarrow 0$.
This is larger than that of any 
element
and large enough for $\delta-$Pu$_{0.95}$Al$_{0.05}$
to be classed as a heavy-fermion system. 
By contrast, $\gamma = 17 \pm 1$~mJK$^{-2}$mol$^{-1}$
in $\alpha-$Pu.
Two distinct
anomalies are seen in the electronic 
contribution to the heat capacity of
$\delta-$Pu$_{0.95}$Al$_{0.05}$, one or both
of which may be associated with
the formation of the $\alpha'-$ martensitic
phase.
We suggest that the large $\gamma$-value of 
$\delta-$Pu$_{0.95}$Al$_{0.05}$
may be caused by
proximity to a quantum-critical point.
\end{abstract}


\maketitle
\date{today}
Plutonium represents the boundary between localised (Am) and
delocalised (Np) $5f$ electrons in the Actinide 
series~\cite{ldrd5,ldrd6}; the resultant
small energy scales, large density of states and general instability
of the $5f$-electron system may be
the root cause of many of Pu's extraordinary 
properties~\cite{ldrd5,ldrd6,ldrd1,ldrd2,ldrd3,ldrd4,plutebook}. 
For instance, it
is thought that itinerant $5f$ electrons lower their energy by
causing Peierls-like distortions, yielding the low-temperature
$\alpha$ (monoclinic), $\beta$ (body-centred monoclinic) and
$\gamma$ (body-centred orthorhombic) phases~\cite{ldrd4,ldrd8}. By
contrast, it is believed that some or all of the $5f$ electrons
are localised in the $\delta$
phase, allowing the Madelung potential of the
remaining $s$, $p$ and $d$ electrons to produce a
higher symmetry face-centred cubic 
structure~\cite{ldrd5,ldrd6,ldrd8}.
Very little provocation is required to transform
the low-symmetry phases into $\delta-$Pu;  
the $\delta$ phase occurs between 319 and $451^{\circ}$C in pure Pu
and is stabilised to zero temperature
by adding a small amount of a trivalent
element, such as Al, Ce or Ga~\cite{plutebook}.

A reliable estimate of the
electronic contribution to the entropy
of Pu is a very important key in understanding the difference
between the $\alpha-$ and $\delta-$phases and the
dramatic effect of alloying.
Unfortunately, attempts to extract relevant
information from $C_P$, the experimental heat
capacity~\cite{sandenaw61,sandenaw62,lee65,lee68,
taylor65,gordon76,stewart85,sandenaw60},
have been inconclusive because the 
phonon contribution to $C_P$ was unknown.
A traditional way to circumvent this problem is
to use low-temperature $C_P$ data;
a plot 
of $C_P/T$ versus $T^2$, where $T$ is the temperature,
is linear at sufficiently low $T$~\cite{ashcroft};
\begin{equation}
(C_P/T)=\gamma + \alpha T^2.
\label{eqn1}
\end{equation}
Here, $\gamma T$ and $\alpha T^3=(12 \pi^4 RT^3)/(5 \theta_{\rm D}^3)$
are the electronic and phonon contributions
to $C_P$; $\theta_{\rm D}$ is the Debye temperature~\cite{ashcroft}.
The $T=0$ intercept, $\gamma$,
is a measure of the electronic density of states.
Sadly, most measurements of $C_P$ in Pu have been restricted to
$T~\gtsim ~ 10$~K, due to problems associated with
self-heating caused by radioactive
decay~\cite{sandenaw61,sandenaw62,lee65,lee68,
taylor65,gordon76,stewart85,sandenaw60}.
In spite of some pioneering work down to $T \approx 7$~K
in $\alpha-$Pu~\cite{gordon76} and $T\approx 4$~K in
$\delta-$Pu$_{1-x}$Al$_x$~\cite{stewart85},  there is still a considerable
spread in the $\gamma$ values reported in the literature; {\it e.g.}, in the
low-$T$ $\delta-$Pu$_{1-x}$Al$_x$ measurements~\cite{stewart85} the $\gamma$
values range from 42 to 68~mJK$^{-2}$mol$^{-1}$. 

In this Letter, we report the solution of these
problems by; (i)~measuring $C_P$ for $\alpha$-Pu and Al-stabilised
$\delta$-Pu to significantly lower temperatures than has been
previously possible ($T\approx 2$~K),
using a sample mount which minimises the effect of self-heating;
and (ii)~extracting the electronic component of $C_P$ for
$\delta-$Pu$_{0.95}$Al$_{0.05}$ by subtracting the phonon
contribution, deduced using recent neutron-scattering data
on the same sample, from
the raw data. These procedures show that the electronic
contribution to the heat capacity varies linearly with $T$ only
when $T~\ltsim ~10 $~K. Moreover, we observe two distinct anomalies
in the electronic heat capacity of
$\delta$-Pu$_{0.95}$Al$_{0.05}$, one or both of which may be associated 
with the $\alpha'-$ martensitic phase observed by
optical metallography. By restricting our analysis to suitably low
temperatures, we obtain $\gamma = 64 \pm 3$~mJK$^{-2}$mol$^{-1}$
for $\delta$-Pu$_{0.95}$Al$_{0.05}$ and 
$\gamma = 17 \pm 1$~mJK$^{-2}$mol$^{-1}$ 
for pure $\alpha$-Pu in the limit $T \rightarrow 0$.
We also observe a large difference in
the electronic contribution to the total entropy 
for $\alpha-$Pu and $\delta-$Pu$_{0.95}$Al$_{0.05}$.

The $\alpha$-Pu sample was prepared by levitation zone refining
and distillation as described in Ref.~\cite{pugrowth}.
Starting material was double-electrorefined $^{242}$Pu cast
into rods. The rods were purified by passing a
10~mm-long molten floating zone ($750^{\circ}$C) ten times through
a cast rod at a travel rate of 1.5~cm/h at
$10^{-5}$~Pa~\cite{pugrowth}. After this, the impurity
level was $174 \pm 26$~ppm, of which U forms approximately
110~ppm~\cite{pugrowth}. The $\delta$-Pu specimen was alloyed by
arc melting followed by a lengthy anneal at $450^{\circ}$C. The
specimen was formed into a plate by rolling followed by heat
treatments to relieve the cold work. Samples were cut,
mechanically polished, chemically polished and heat treated prior
to measurement.

Heat capacity measurements were made using the thermal relaxation
method in a Quantum Design PPMS, the performance of which has been
subjected to extensive analysis~\cite{jason}. To
counteract the self-heating due to radioactive decay, a modified
sample puck with high thermal contact to the heat bath 
was employed for the low-$T$ data.
Measurements comparing the modified puck with a
standard one at higher $T$ were identical within experimental
error. Measurements made from 10~K to 300~K used
samples ranging from 20 to 30~mg, while below 10~K, sample
masses of
5 to 10~mg were used. 
Samples were secured to the puck using
Apiezon N-grease to ensure good thermal contact. Immediately
before each sample was studied, the addenda (puck and grease) were
measured over the same $T$ range. All heat capacities
shown in the figures are corrected by subtracting the addenda
contribution from the raw data; systematic errors (shown as bars)
due to inaccuracies in the PPMS~\cite{jason} and measurement
of the sample masses are $\approx \pm 1.5$\% of $C_P$.

\begin{figure}
\centering
\includegraphics[height=8.5cm]{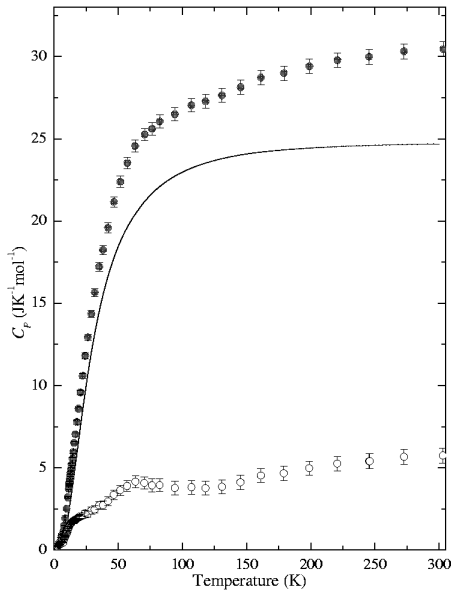}
\centering \caption{Experimental heat capacity
of $\delta-$Pu$_{0.95}$Al$_{0.05}$ (filled points: $\bullet$)
versus temperature.
The curve is $C_{P{\rm ph}}$,
the phonon contribution to the heat capacity.
The electronic contribution
to the heat capacity ($C_{\rm el}=C_P-C_{P{\rm ph}}$)
is plotted as hollow points ($\circ$).
}
\label{fig1}
\end{figure}

The heat capacity $C_P$ of $\delta$-Pu$_{0.95}$Al$_{0.05}$
is plotted versus $T$ in Fig.~\ref{fig1}
(solid points).
To extract the electronic contribution to $C_P$,
we employ a recent measurement of the phonon
density of states $g(E)$ as a function of energy $E$
carried out on the same sample
of $\delta$-Pu$_{0.95}$Al$_{0.05}$~\cite{neutron}.
Neutron-scattering and sound-velocity data were
used to derive $g(E)$ at $T=27$, 65, 150 and
300~K~\cite{neutron}.
The phonon contribution to
$C_P$, $C'_{P{\rm ph}}$,
was computed using
\begin{equation}
C'_{P{\rm ph}} \approx C_{V{\rm ph}}=\frac{\partial~}{\partial T}(
\int^{\infty}_0 E g(E) f(E,T) {\rm d}E ).
\end{equation}
Here $C_{V{\rm ph}}$ is the phonon heat capacity at constant
volume~\cite{correction}, $E$ is the energy
and $f(E,T)$ is the Bose-Einstein distribution function.

Such an approximation neglects anharmonic effects;
however, the $T-$dependence of $g(E)$~\cite{neutron}
shows that such effects are small for $T~ \ltsim ~150$~K.
More significantly, the computed
$C'_{P{\rm ph}}$ varied by up to $\pm 1$\% (i.e. of similar
size to the experimental uncertainty in $C_P$), depending
on which $g(E)$ (i.e. that based on the 27, 65, 150 or 300~K data)
was used. To minimise the impact of this effect,
the phonon contribution to the heat capacity
$C_{P{\rm ph}}$ plotted in Fig.~\ref{fig1} (curve) is a
$T-$dependent interpolation
between the computed $C'_{P{\rm ph}}$.

\begin{figure}
\centering
\includegraphics[height=8.5cm]{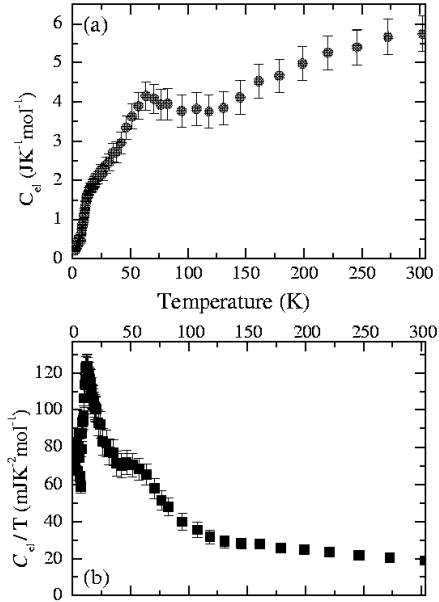}
\centering \caption{(a) Electronic contribution to
the heat capacity of
of $\delta-$Pu$_{0.95}$Al$_{0.05}$
($C_{\rm el}=C_P-C_{P{\rm ph}}$) versus $T$.
(b)~The same data plotted as
$C_{\rm el}/T$ versus $T$.
} \label{fig2}
\end{figure}

$C_{\rm el}$, the electronic contribution to $C_P$ of
$\delta$-Pu$_{0.95}$Al$_{0.05}$, was estimated by subtracting
$C_{P{\rm ph}}$ from the experimental $C_P$
data~\cite{correction}; $C_{\rm el}$ values are shown as hollow
points in Fig.~\ref{fig1} and on an expanded vertical scale in
Fig.~\ref{fig2}a. As noted in the discussion of
Eq.~\ref{eqn1}, the expectation for a simple metal is that
$C_{\rm el}=\gamma T$. Even a cursory inspection of
Fig.~\ref{fig2}a shows that the experimental values
of $C_{\rm el}$ only follow a straight line through the origin for
 $T~\ltsim 10$~K; between approximately 10 and 40~K,
there is a distinct ``hump'' superimposed on the quasilinear
increase, whilst at $T \approx 65$~K there is a
``$\lambda$-shaped'' maximum, eventually followed by a more gentle
increase. 

\begin{figure}
\centering
\includegraphics[height=4.0cm]{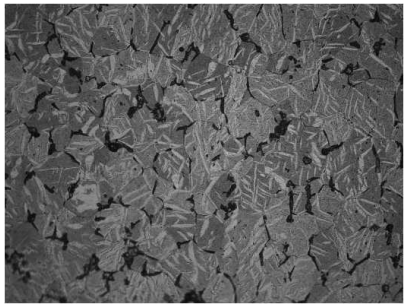}
\centering \caption{Optical metallography showing the
surface of the $\delta-$Pu$_0.95$Al$_0.05$ heat capacity
sample after the
measurement.
The $\alpha '$ martensite phase is identified as the
light ``tweed'' pattern on the
surface. The sample was photographed at
$500\times$, and the standard ASTM method was
used to determine a 5 - 10 \% volume fraction of
the martensite (light acicular formations).
} \label{fig2c}
\end{figure}

A $\lambda-$like feature in the heat capacity
is characteristic of a martensitic transition~\cite{martensite}.
Support for this attribution comes from
the retention of a small
fraction of the $\alpha'-$ phase, 
as revealed by the characteristic
``tweed'' structure shown in a metallographic examination of the
sample after thermal cycling (Fig.\ref{fig2c}).
Neutron-scattering and elastic-moduli data on the same sample
before and after cooling~\cite{neutron}, and
volume-fraction analysis of optical metallography suggest
that our $\delta$-Pu$_{0.95}$Al$_{0.05}$ contains around $5-7$\%
of the $\alpha'-$ phase. 
Note that a knowledge
of the phonon contribution was required to reveal the
martensite feature in the heat capacity; 
until the current work,
no clear indication of such a phase
has been extracted from the heat capacity of $\delta-$Pu.
Moreover, the manifestation of the transition 
in $C_{\rm el}$ strongly suggests that the transition 
is electronically driven. 

Fig.~\ref{fig2}b shows the effective $\gamma~(=C_{\rm el}/T)$
for $\delta$-Pu$_{0.95}$Al$_{0.05}$, plotted as a function
of $T$. For $T~\ltsim 10$~K,
$\gamma \approx 65$~mJK$^{-2}$mol$^{-1}$.
Around 10~K, there is a sharp dip,
followed immediately by the above-mentioned ``hump'' in $C_{\rm el}$,
which appears as a broad peak (maximum at $T\approx 13$~K)
in the effective $\gamma$.
Such a peak suggests a contribution to the
electronic entropy associated with a second
phase transition at $T\approx 13$~K.
This may be linked to the $\lambda$-like
transition seen in $C_{\rm el}$ at $T\approx 65$~K (Fig.~\ref{fig2}a);
multistage phase transitions have been 
observed in
actinides such as U and predicted in Pu~\cite{saxena}.

Above 40~K, $C_{\rm el}/T$ returns briefly to
$\gamma \approx 70$~mJK$^{-2}$mol$^{-1}$,
before falling gradually to $\gamma \approx 20$~mJK$^{-2}$mol$^{-1}$.
This complicated variation illustrates the great importance
of low-temperature (i.e., $T~\ltsim ~10$~K) $C_P$ data.
The non-linear variation of the
electronic contribution to the heat capacity
with $T$ is the probable reason for the previous,
widely-varying values of $\gamma$ and $\theta_{\rm D}$
for $\delta-$Pu
quoted in the literature~\cite{taylor65,stewart85,sandenaw60}.

\begin{figure}
\centering
\includegraphics[height=8.5cm]{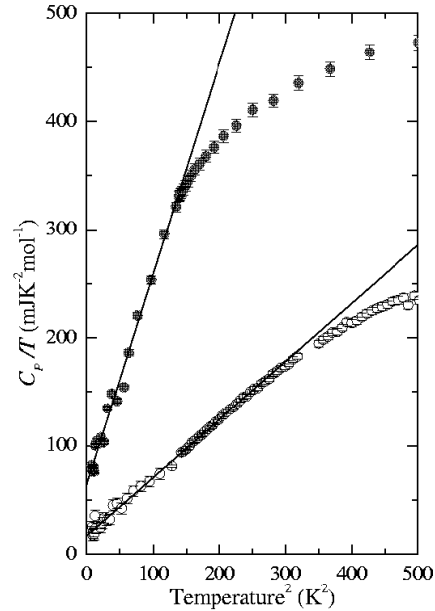}
\centering \caption{Low-temperature
values of $C_P/T$ for both the
pure $\alpha-$Pu ($\circ$)
and $\delta-$Pu$_{0.95}$Al$_{0.05}$ ($\bullet$)
samples plotted as a function of $T^2$;
the low-$T$
portions of the data have been fitted to Eq.~\ref{eqn1}.
}
\label{fig3}
\end{figure}

Having established that the electronic
contribution to the heat capacity of
$\delta-$Pu$_{0.95}$Al$_{0.05}$ is linear in
$T$ only below $T \approx 10$~K,
we perform a fit of Eq.~\ref{eqn1} to the
experimental $C_P$ data in this range;
this is shown $(\bullet)$
in Fig.\ref{fig3}
which also displays
$C_P/T$ for pure $\alpha-$Pu~$(\circ)$.
Similarly, the fit for $\alpha-$Pu
is restricted to $T< 16$~K.
The fits of Eq.~\ref{eqn1} yield
$\gamma = 64 \pm 3$~mJK$^{-2}$mol$^{-1}$
(in good agreement with Fig.~\ref{fig2}b, and lying within the spread
of values reported in Ref.~\cite{stewart85})
and $\theta_{\rm D}=100 \pm 2$~K
for $\delta-$Pu$_{0.95}$Al$_{0.05}$.
Likewise, we obtain
$\gamma = 17 \pm 1$~mJK$^{-2}$mol$^{-1}$
({\it i.e.} within the  range $16-23$~mJK$^{-2}$mol$^{-1}$ 
reported by Ref.~\cite{gordon76})
and $\theta_{\rm D}=153 \pm 2$~K for
$\alpha-$Pu.

The value of $\gamma$ for $\alpha-$Pu is 
remarkable enough, being bigger than
that of any other element~\cite{lee68,phillips};
nevertheless its large size may be understood
reasonably conventionally when the $5f$ electrons
are taken into account~\cite{lee68}.
However, $\gamma$ for $\delta-$Pu$_{0.95}$Al$_{0.05}$
is a factor $\sim 4$ bigger,
being large enough to class it 
as a heavy-fermion system~\cite{hewson}.
Note that the increase cannot be simply related to
the presence of Al, which has a comparitively small value
of $\gamma$ in its pure form~\cite{phillips}.

Finally, we compute the specific entropies using
\begin{equation}
S_{\rm el}=\int^{300}_0 \frac{C_{\rm el}}{T}{\rm d}T~~~{\rm and}~~~~
S_{\rm total}=\int^{300}_0 \frac{C_P}{T}{\rm d}T.
\end{equation}
For $\delta-$Pu$_{0.95}$Al$_{0.05}$, we
find that $S_{\rm el}=11.4$~JK$^{-1}$mol$^{-1}$,
of which approximately 2~JK$^{-1}$mol$^{-1}$ is associated
with the peak in $C_{\rm el}/T$ at $T\approx 13$~K;
this should be compared with $S_{\rm total}=68.4$~JK$^{-1}$mol$^{-1}$.
By contrast, $S_{\rm total}=57.1$~JK$^{-1}$mol$^{-1}$
for $\alpha-$Pu. Although the lack of neutron data
means that we do not have a reliable
means of extracting $C_{\rm el}$ in $\alpha-$Pu,
an upper bound for $S_{\rm el}$ is given by
$300 \times \gamma \approx 5.1$~JK$^{-1}$mol$^{-1}$.
Hence $S_{\rm el}/S_{\rm total}~ \ltsim ~0.09$ for
$\alpha-$Pu, roughly half the value 
$S_{\rm el}/S_{\rm total} \approx 0.17$ 
obtained for $\delta-$Pu$_{0.95}$Al$_{0.05}$.
As in the case of $\gamma$,
the $S_{\rm el}/S_{\rm total}$ values suggest that
the role of the electronic system
is enhanced on
going from the $\alpha-$ to 
the $\delta-$ phase.

In some respects, the behavior of
Pu is similar to models of quantum
criticality~\cite{ldrd16,ldrd18} which 
associate  quantum-critical points with
rearrangements of the Fermi surface,
due either to charge- or spin-density-wave-like
reconstruction (analogous to
the Peierls-like distortions thought to
occur in the $\alpha-$phase~\cite{ldrd8}), 
or to the onset of itineracy
for previously localised electrons (as may occur in
the transition from $\delta-$ to $\gamma-$Pu~\cite{ldrd8}).
A characteristic feature of a quantum-critical point
is the proximity of many excited states to the groundstate,
consistent with the anomalously large (for an element) value
of $\gamma$ seen in $\delta-$Pu~\cite{ldrd16}.
All of the strange properties of Pu,
including the complex phase diagram, 
may, in fact, be the result of $\delta-$Pu
being close to a quantum-critical point. 
This could imply that the
properties of Pu are ``emergent'', and not easily derivable 
from microscopic models.

In summary,
we have measured the heat capacities of $\delta-$Pu$_{0.95}$Al$_{0.05}$
and $\alpha-$Pu over the temperature range $2-303$~K.
The availability of data below 10~K plus an
estimate of the phonon contribution to the heat capacity based on
neutron-scattering data enable us to make a
reliable deduction of the low-temperature electronic
contribution to the heat capacity of $\delta-$Pu$_{0.95}$Al$_{0.05}$;
we find $\gamma = 64 \pm 3 $~JK$^{-2}$mol$^{-1}$.
By contrast, $\gamma = 17 \pm 1$~JK$^{-2}$mol$^{-1}$
in $\alpha-$Pu.
We note two anomalies in the electronic contribution
to the heat capacity of $\delta-$Pu$_{0.95}$Al$_{0.05}$,
one or both of which may be 
associated with a martensitic phase transition.
The large increase in $\gamma$ and the electronic contribution
to the entropy on going from $\alpha-$ to $\delta-$Pu may
be associated with the proximity of $\delta-$Pu
to a quantum-critical point.

This work is supported by the US
Department of Energy (DoE) under
grant LDRD-DR 20030084 ``Quasiparticles and phase transitions in high
magnetic fields:  critical tests of our understanding of Plutonium''.
Part of this work is carried out at the National High Magnetic Field Laboratory,
which is supported by the National Science Foundation, the
State of Florida and DOE.
We thank Fivos Drymiotis, Mike Ramos and Ramiro
Pereyra  for experimental assistance and
George Chapline, Neil Harrison and Chuck Mielke
for interesting comments.


\begin{thebibliography}{Bibliography}
\bibitem{correction}
In the analysis,
we neglect the constant volume to
constant pressure correction for the
heat capacity, $C_P-C_V =\beta^2 \nu B T$,
where $\beta$ is the volume expansion coefficient,
$\nu$ is the specific volume, and $B$ is the bulk
modulus~\cite{thermobook}.
Using standard values~\cite{plutebook},
$\beta^2 \nu B T$ is $\ltsim 0.25$\% of $C_P$,
at $T=150$~K,
and therefore smaller than typical experimental errors.
\end{thebibliography}
\end{document}